\documentclass[reprint,
 amsmath,amssymb,
 aps,
floatfix,
showkeys
]{revtex4-1}

\usepackage[USenglish]{babel}
\usepackage{slashed}
\usepackage{physics}
\usepackage{graphicx}% Include figure files
\usepackage{dcolumn}% Align table columns on decimal point
\usepackage{bm}% bold math
\usepackage[usenames]{color}
\usepackage{xcolor}
\usepackage{float}

\newcommand{\be}{\begin{equation}}
\newcommand{\ee}{\end{equation}}
\newcommand{\bea}{\begin{eqnarray}}
\newcommand{\eea}{\end{eqnarray}}
\newcommand{\beas}{\begin{eqnarray*}}
\newcommand{\eeas}{\end{eqnarray*}}

\newcommand{\plog}[1]{{\text{Li}}_{#1}}
\begin{document}

\title{Thermomagnetic renormalons in a scalar self-interacting $\lambda \phi^4$ theory}% Force line breaks with \\


\author{M. Loewe$^{1,2,3}$, L. Monje$^{1}$, and R. Zamora$^{4,5}$ }
\affiliation{%
$^1$Instituto de F\'isica, Pontificia Universidad Cat\'olica de Chile, Casilla 306, Santiago 22,Chile.\\
$^2$Centre for Theoretical and Mathematical Physics, and Department of Physics, University of Cape Town, Rondebosch 7700, South Africa.\\
$^3$Centro Cient\'ifico-Tecnol\'ogico de Valpara\'iso CCTVAL, Universidad T\'ecnica Federico Santa Mar\'ia, Casilla 110-V, Valpara\'iso, Chile.\\
$^4$Instituto de Ciencias B\'asicas, Universidad Diego Portales, Casilla 298-V, Santiago, Chile.\\
$^5$Centro de Investigaci\'on y Desarrollo en Ciencias Aeroespaciales (CIDCA), Fuerza A\'erea de Chile, Casilla 8020744, Santiago, Chile.}%


\begin{abstract}
In this article we extend a previous discussion about the influence of an external magnetic field on renormalons in a self interacting scalar theory by including now temperature effects, in the imaginary formalism, together with an external weak external magnetic field. We show that the location of poles in the Borel plane does not change, getting their residues, however, a dependence on temperature and on the magnetic field.The effects of temperature and the magnetic field strength on the residues turn out to be opposite. We present a detailed discussion about the evolution of these residues, showing  technical details involved in the calculation.   

\end{abstract}

%\pacs{}% PACS, the Physics and Astronomy
\keywords{Renormalons, Effective Models, Magnetic Fields}%Use showkeys class option if keyword

\maketitle

%%%%%%%%%%%%%%%%%%%%%%%%%%%%%%%%%%%%%%%%%%%%%%%%%%%%%%%%%%%%%%%%%
\section{Introduction}\label{sec1}
%%%%%%%%%%%%%%%%%%%%%%%%%%%%%%%%%%%%%%%%%%%%%%%%%%%%%%%%%%%%%%%%%
Since the paper by Dyson \cite{Dyson} on the convergence of perturbative series in quantum electrodynamics (QED), a seminal article based exclusively on physical arguments, we have learned that power series expansions in Quantum Field Theory (QFT), in general, are divergent objects. For high orders  of perturbation, the divergence grows like $n!$, where n is the order of expansion, and this is due, essentially, to the multiplicity of diagrams that contribute to a certain Green function, or to a physical process, at such expansion order. A usual procedure to improve the convergence relies on the Borel transform \cite{Altarelli,Rivasseau,Khanna}. However, in some cases, even the Borel transformed series are divergent, spoiling  the meaning of the whole procedure. The new singularities responsible for this divergent behavior are the renormalons. For a review, see \cite{Beneke}. Recently there has been a renewed interest in the subject by considering one loop renormalization group equation  in multi-field  theories \cite{Vasquez} or by considering finite temperature mass correction in the $\lambda \phi ^{4}$ theory, reanalyzing the temperature dependence of poles and their residues \cite{Cavalcanti}. There are other sources of divergences, as instantons in quantum chromodynamics (QCD) \cite{Gross,Shafer}, for example. However, we will not refer here to such objects that can be handled using semi-classical methods. In peripheral heavy ion collisions, huge magnetic fields appear \cite{warringa}. In fact, the biggest fields existing in nature. The interaction between the produced pions in those collisions may be strongly affected by the magnetic field.Temperature is, of  course, also present in such scenario. In fact, several studies on the behavior of different physical parameters in the presence of external magnetic field and/or temperature have been carried on by different authors. \cite{bali01,iranianos,zamora1,zamora2,zamora3,simonov03,aguirre02,tetsuya,dudal04,kevin,gubler,noronha01,morita,Ayala0,Ayala1,morita02,sarkar03,band,nosso1,nosso03,Ayala2,Ayala3,zamora4}. In this article we analyze, in the frame of a self interacting scalar  $\lambda \phi^{4}$ theory, the influence of the magnetic field and temperature on the position of the U.V. renormalons (the only relevant in $\lambda\phi^4$ theory) and their residues in the Borel plane. Recently, some extensions as the q-Borel series have been proposed, allowing the discussion of series whose coefficients grow like $(k!)^q$ \cite{Cavalcanti2}. Also, there have been new attempts to find corrections to the Beta function in QED with $N_{f}$ flavors by considering closed chains of diagrams, like renormalons, and computing corrections of order $N _{f}^{-2}$ and $N_{f}^{-3}$ \cite{Dunne}. These authors found a new logarithmic branch cut whose physical role is not clear. Probably the same situation will occur in self interacting scalar theories with several components. In the present article we extend previous discussions where the behavior of renormalons in $\lambda \phi ^4$ was analyzed, separately, for the case where thermal corrections appeared in the real time formalism, and for corrections due to the presence of a  magnetic field. Here we present a discussion where both effects are simultaneously taken into account. As we will see, we need to handle with care the different possible scenarios. We discuss the high and low temperature cases for the weak magnetic field region, identifying the temperature and magnetic field dependent sub-leading poles in the Borel plane. In the strong magnetic field regime there are no new singularities. A physical discussion will be presented. This article is organized as follows. In section II we go through general aspects concerning a temperature and magnetic field dependent scalar propagator. In section III we present a brief discussion about the concept of Borel summable series and poles in the Borel plane. In Section IV we present the pure thermal discussion in the imaginary time formalism. Section V presents the general  discussion, including temperature and magnetic field effects, about renormalon corrections to the propagator. Some technical details can be found in an appendix. Finally we present our conclusions.

%%%%%%%%%%%%%%%%%%%%%%%%%%%%%%%%%%%%%%%%%%%%%%%%%%%%%%%%%%%%%%%%%%%%

%naThe discussion of the thermal evolution of $\pi$-$\pi$ scattering
%lengths turns out to be a relevant problem in the context of heavy
%ion collisions. In fact, we know that in the central rapidity
%region, precisely where the quark-gluon plasma is expected to be
%created, a big amount of thermalized pions will be produced. Those
%pions will interact among themselves and the $\pi$-$\pi$
%scattering lengths are crucial parameters in order to describe the
%scattering amplitudes. At zero temperature the $\pi$-$\pi$
%scattering lengths were first measured by Rosellet et al
%\cite{Rosellet}. A recent review about the present status of the
%experimental measurements of $\pi$-$\pi$ scattering lengths and
%their comparison with different
%theoretical approaches can be found in \cite{Urets}.
%In this article we will present a detailed calculation of the
%%thermal corrections to the $\pi$-$\pi$ scattering lengths in the
%frame of the linear sigma model \cite{Gell-Mann}. As it is well
%known, the linear sigma model is an effective, renormalizable
%\cite{Lee}, low energy description of hadron dynamics. Our
%calculations will be done using the real time
%formalism at the one loop level.

%%%%%%%%%%%%%%%%%%%%%%%%%%%%%%%%%%%%%%%%%%%%%%%%%%%%%%%%%%%%%%%%%%%%
\section{Thermomagnetic renormalon-type correction to the propagator} \label{magneticpropagator}
Thermomagnetic corrections will be handled through Schwinger's bosonic propagator at finite temperature. We are going to introduce the propagator taking first only the external magnetic field into account, incorporating then later finite temperature effects.   Schwinger's bosonic propagator is given by \cite{Schwinger}
\begin{equation} \label{propfase}
D^B(x',x'')=\phi(x',x'')\int \frac{d^4k}{(2\pi)^4} e^{-ik\cdot(x'-x'')}D^B(k) \text{,}
\end{equation}
where
\begin{eqnarray}\label{prop1}
iD^B(k)&=&\int_0^\infty \frac{ds}{\cos(eBs)} \nonumber \\
&\times&\exp \left( is \left[ k_\parallel^2-k_\perp^2\frac{\tan (eBs)}{eBs}-m^2+i\epsilon \right] \right).
\end{eqnarray}

The 4-momentum has been decomposed into parallel and perpendicular components respect to the magnetic field direction. By considering a constant magnetic field, whose direction defines the z-axis, we can write
\begin{equation}
(a \cdot b)_\parallel = a^0b^0 -a^3b^3 , \qquad (a \cdot b)_\perp = a^1b^1+a^2b^2,
\end{equation}
for two arbitrary four vectors $a_{\mu}$ and $b_{\mu}$. We have also
\begin{equation}
a^2 = a_\parallel ^2 - a_\perp ^2.
\end{equation}

Notice that the phase factor in Eq. (\ref{propfase}), given by
\begin{equation}\label{fase}
\phi(x',x'')=\exp \left(  ie \int_{x''}^{x'} dx_\mu A^\mu (x)  \right) ,
\end{equation}
which is independent of the path, can be ignored for two-point functions which are diagonal in configuration space, as it is the case in our analysis. As usual, we shall work in the momentum representation.
In this way using $eBs \rightarrow s$ we find
\begin{eqnarray} \label{propmin}
&&iD^B(k)=\frac{1}{eB}\int_0^\infty \frac{ds}{\cos(s)} \nonumber \\
&&\times \exp \left( i(s/eB) \left[ k_\parallel^2-k_\perp^2\frac{\tan (s)}{s}-m^2+i\epsilon \right] \right).
\end{eqnarray}
This propagator can be expressed as a sum over Landau levels \cite{Ayalaetal}
\begin{equation}\label{proplandau}
iD^B(k)=2i \sum_{l=0}^\infty \frac{(-1)^l L_l \left( \frac{2k_\perp^2}{eB} \right) e^{-k_\perp^2/eB}}{k_\parallel^2-(2l+1)eB - m^2 +i\epsilon},
\end{equation}
where $L_l$ are the Laguerre polynomials. By considering the previous expression in the region where $eB \ll m^{2}$ it is possible to show  that \cite{gluon}
\begin{eqnarray}\label{propmagneticodebil}
iD^B(k)&&\xrightarrow{eB\rightarrow 0} \frac{i}{k_\parallel^2 - k_\perp^2 -m^2} - \frac{i(eB)^2}{(k_\parallel^2 - k_\perp^2 -m^2)^3} \nonumber \\
&-& \frac{2i(eB)^2k_\perp^2}{(k_\parallel^2 - k_\perp^2 -m^2)^4},
\end{eqnarray}
which is the desired weak field expansion for our calculation. It is straightforward to generalize these expressions to the finite temperature scenario through an analytic continuation into Matsubara frequency space \cite{LeBellac, Kapusta}, i.e. 
\begin{equation}
k_0 \rightarrow i\omega_{n} = \frac{2\pi n}{\beta},\; n\in Z,
\end{equation} 
where $\beta=1/T$, and where the integral in $k_0$ converts into a sum according to,
\begin{eqnarray}
\int \frac{d^4k}{(2\pi)^4}f(k) \rightarrow \frac{i}{\beta} \sum_{n\in Z} \int \frac{d^3k}{(2\pi)^3} f(i\omega_n,\vec{k}).
\end{eqnarray}
\section{The Borel Transform} \label{Boreltransform}

We will briefly present the Borel transform method, which can be considered as a tool designed to make sense of potentially divergent series \cite{Altarelli}.
\\
For a divergent perturbative expansion 

\begin{equation}
D[a]=\sum_{n=1}^\infty D_na^n \text{,}
\end{equation}

\noindent
the Borel transform $B[b]$ of the series $D[a]$ is defined through 

\begin{equation}
B[b]=\sum_{n=0}^\infty D_{n+1}\frac{b^n}{n!} \text{.}
\end{equation}

The inverse transform is, 

\begin{equation}
D[a]=\int_0^\infty db \hspace{2mm} e^{-b/a} B[b] \text{.}
\end{equation}

We need the last integral to be convergent, in order to make sense to the series, being $B[b]$ free form singularities in the range of integration. If these conditions are fulfilled, we say that the original series $D[a]$ is Borel summable.

It is easy to check that all convergent series are also Borel summable. For the case of divergent series this is not necessary the case. If we find poles in the $0-\infty $ range of integration of the previous equation, the series is no longer Borel summable. It is possible, however, to make sense to this integral, in such cases, through a prescription for the integration path in the complex $[b]$-plane, avoiding the pole. This will be, however, a prescription-dependent result.

A classical reference about divergent series is the book by Hardy, \cite{hardy}.

It is known that perturbative expansions in quantum field theory are not Borel summable. There are two sources for the appearance of singularities in the Borel plane: renormalons and instantons. Here we do not want to comment about the latter possibility. In QCD, renormalons have been a matter of debate since these objects may affect our understanding of the gluon condensate \cite{Beneke}.

\section{Renormalons in the vacuum}\label{secRenVacio}
 In the $\lambda \phi^{4}$ theory the renormalon type diagrams that produce poles in the Borel plane correspond to a correction to the two-point function

\vspace{0.5cm}
\begin{figure}[h] 
\begin{center} 
\includegraphics[width=7cm]{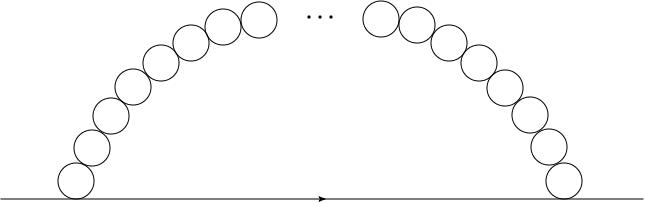}
\caption{Renormalon-type contribution to the two point function.}
\label{fig:diagrenormalon}
\end{center}
\end{figure}

First we revise the calculation of this diagram with the insertion of k bubbles, summing then over $k$ and studying the behavior of its transform in the Borel plane.

We will denote by $R_k(p)$ the diagram of order $k$ shown in Fig. {\ref{fig:diagrenormalon}}, where  $p$ is the four-momentum  entering and leaving the diagram and  $q$ is the four momentum that goes around the chain of bubbles

\begin{equation}\label{Rk}
R_k(p)=\int \frac{d^4q}{(2 \pi)^4}\frac{i}{(p+q)^2-m^2+i\epsilon}\frac{[B(q)]^{k-1}}{(-i\lambda)^{k-2}} \text{.}
\end{equation}

\begin{figure}[h]
\begin{center}
\includegraphics[scale=0.5]{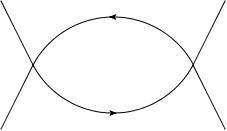} 
\caption{Fish-diagram.}
\label{fig:diagfish}
\end{center}
\end{figure}

In this expression,  $B(q)$ corresponds to the contribution of one bubble in the chain which is equivalent, of course, to the one-loop correction, in the s-channel, of the vertex, the so called   fish-diagram  (see Fig. \ref{fig:diagfish}). The factor $(-i\lambda)^{k-2}$ has been added to cancel the vertices that have been counted twice along the chain.

The expression for  $B(q)$ is a well known result \cite{bailin} 
\begin{equation}
\begin{split}
B(q)=\frac{(-i\lambda)^2}{2}\int  \frac{d^4k}{(2 \pi)^4}\frac{i}{k^2-m^2+i\epsilon} \\ \times\frac{i}{(k+q)^2-m^2+i\epsilon}
\end{split}
\end{equation}
\noindent given by
\begin{equation}\label{fishexact}
\begin{split}
B(q)=\frac{-i\lambda^2}{32\pi ^2} \left[ \Delta  - \int _0 ^1 dx \log \left( \frac{m^2 - q^2 x(1-x) - i\epsilon}{\mu ^2} \right) \right] 
\end{split}
\end{equation}

\noindent where $\Delta$ is the divergent part that will be canceled by the counterterms, being $\mu$ an arbitrary mass-scale  associated to the regularization procedure,  which always appear when we go through the renormalization program.

The contribution to the renormalon comes from the deep euclidean region in the integral in Eq.( \ref{Rk}), i.e. where  $-q^2\gg m^2$. In this way,  $B(q)$ in Eq. (\ref{fishexact}) can be approximated as:

\begin{equation}\label{fishDE}
B(q) \approx \frac{-i\lambda^2}{32\pi ^2} \log(-q^2/\mu^2) \text{,}
\end{equation}

Inserting this result in Eq. (\ref{Rk}) and performing a Wick rotation, we find

\begin{eqnarray}
R_k(p)&=&\frac{-i\lambda^k}{(32\pi ^2)^{k-1}} \int \frac{d^4q}{(2 \pi)^4}\frac{1}{(p+q)^2+m^2} \nonumber \\
&\times& [\log(q^2/\mu^2)]^{k-1} \text{.}
\end{eqnarray}

This is an ultraviolet divergent expression. However, since the theory is renormalizable, we can separate this expression into  a finite and a divergent part. We are only interested in the finite part. For this we expand the propagator in powers of $1/q^2$, neglecting the first two terms that leave divergent integrals. As we know this is justified due to the appearance of the counterterms.

So, we find
\begin{align}\label{R2}
\begin{split}
R_k =& -i \left( \frac{\lambda}{32\pi ^2} \right)^k 4 m^4 \int _\Lambda ^\infty dq [\log (q^2/\mu^2)]^{k-1}q^3 \\
&\times \left[\frac{m^4}{q^6}-\frac{m^6}{q^8}+...\right]
\end{split}
\end{align}
with $\Lambda >0$.
The dependence on the external momentum $p$ has disappeared. The lower limit $\Lambda $ comes form the fact that we are interested in the deep euclidean region. It has to be fixed in order to fulfill this condition.

Introducing $q=\mu e^t$ in the first two terms of Eq. (\ref{R2}), and fixing $\Lambda = \mu$, we find

\begin{align}
R_k =&  -i \left( \frac{\lambda}{32\pi ^2} \right)^k  \frac{4 m^4}{\mu^2}\int_0 ^\infty dt e^{-2t} (2t)^{k-1} \nonumber\\
&\times \left(1-\frac{m^2}{\mu^2} e^{-2t} \right)  \nonumber \\
=&  -i \left( \frac{\lambda}{32\pi ^2} \right)^k \frac{2 m^4}{\mu^2} \Gamma (k) +i \left( \frac{\lambda}{64\pi ^2} \right)^k \frac{2 m^6}{\mu^4} \Gamma (k) \nonumber\\
=& -i \left( \frac{\lambda}{32\pi ^2} \right)^k \widetilde{m}^2 \Gamma (k) +i \left( \frac{\lambda}{64\pi ^2} \right)^k \frac{m^2}{\mu^2}\widetilde{m}^2 \Gamma (k)
\end{align}
with $\widetilde{m}^2=2m^4/\mu^2$. We see that this diagram grows like $k!$, inducing then a pole in the Borel plane.\\

By taking the series $\Sigma R_k$,

\begin{equation}
D[\lambda] = \sum_k \frac{-i}{(32\pi ^2)^k}\Gamma(k)\lambda^k \text{,}
\end{equation}
its Borel transform is given by 
\begin{eqnarray}
B[b] &=& \sum_k \left( \frac{R_k}{\lambda^k} \right)\frac{b^{k-1}}{(k-1)!} \text{,}\nonumber \\
&=& -i \widetilde{m}^2\frac{1}{1-b/32\pi^2} +i \frac{m^2}{\mu^2}\widetilde{m}^2\frac{1}{1-b/64\pi^2} \text{,}
\end{eqnarray}

identifying, finally, the leading pole on the positive semi real axis in the Borel plane $b=32\pi^2$ and the second pole in $b=64\pi^2$.

\section{THERMAL
RENORMALONS}\label{thermorenormalons}
First we are going to calculate the renormalon contribution at finite temperature and in the next section we will consider  both temperature as well as the presence of a weak magnetic field. Although the finite temperature calculation was considered in \cite{Loewe}, this analysis was carried on in the frame of the real time formalism, thermo-field dynamics. We are now going to calculate first the finite temperature renormalon in the imaginary time formalism. There are some interesting technical details, worthwhile to be mentioned. The diagram to be calculated is
\begin{equation}
R_{T,k}=\frac{1}{(-i\lambda)^{k-2}} \int \frac{d^4q}{(2\pi)^4} i D(p+q)[B^T(q)]^{k-1},
\end{equation}
Notice that the $D(p+q)$ propagator could be taken also as temperature dependent. However, in our previous work \cite{malaquias} for the pure magnetic corrections case, we showed that if the propagator $D(p+q)$ included magnetic field contributions, taking also magnetic corrections to  $B(q)$, sub-leading terms appear. Also we showed that if we take now magnetic corrections for $D(p+q)$, being $B(q)$ independent of the magnetic field, it has the same effect as taking $D(p+q)$ without magnetic field and $B(q)$ with magnetic field. The same situation happens also here, when dealing only with temperature corrections. Therefore, our calculation will be carried on by considering $B(q)$ as temperature dependent, whereas $D(p+q)$ will be handled without temperature effects. First let us consider one bubble 
\begin{equation}
B^T(q) =  \frac{(-i\lambda)^2}{2}  i T \sum_{n} \int \frac{d^3k}{(2\pi)^3}   D^T(k) D^T(k-q), \label{fishT}
\end{equation}
where 
\begin{equation}
  D^T(k) = \frac{1}{\omega_n^2+k^2+m^2},  
\end{equation}
therefore
\begin{eqnarray}\label{26}
B^T(q) &=&  \frac{(-i\lambda)^2}{2} i T\sum_{n} \int \frac{d^3k}{(2\pi)^3} \frac{1}{\omega_n^2+k^2+m^2} \nonumber \\
&\times&\frac{1}{(\omega_n-\omega)^2+(k+q)^2+m^2}. 
\end{eqnarray}
The sum over Matsubara frequencies is calculated using well knowing techniques \cite{LeBellac}, getting
\begin{align}
\begin{split}
B^{T}(q) =  \frac{(-i\lambda)^2}{2} i\int & \frac{d^3k}{(2\pi)^3} \frac{- s_1 s_2}{4 E_1 E_2} \\ &\times\frac{1+n(s_1 E_1)+n(s_2 E_2)}{i \omega -s_1 E_1 - s_2 E_2}, 
\end{split}
\end{align}
where 
\begin{center}
\begin{tabular}{lr}
$E_1^2=k^2+m^2$, 		& $s_1=\pm1$, \\
$E_2^2=(k+q)^2+m^2$, 	& $s_2=\pm1$, \\
\multicolumn{2}{c}{$n_i(E_i) = 1/(e^{E_i/T}-1)$}.\\
\end{tabular}
\end{center}

Writing explicitly $s_1$ and $s_2$, we obtain
\begin{align}
B^{T}(q) =&  \frac{(-i\lambda)^2}{2}i \int \frac{d^3k}{(2\pi)^3} \frac{1}{4 E_1 E_2} \nonumber\\
&\times\Biggl[ (1+n_1+n_2) \Biggl( \frac{1}{i\omega - E_1 - E_2} -\frac{1}{i\omega + E_1 + E_2} \Biggr) \nonumber\\  
&-(n_1-n_2)\Biggl( \frac{1}{i\omega - E_1 + E_2} -\frac{1}{i\omega + E_1 - E_2} \Biggr)\Biggr] \nonumber\\
\equiv& B_{\text{vac}}(q)+ B_T(q),
\end{align}
where
\begin{align}
\begin{split}
B_{\text{vac}}(q) =&  \frac{(-i\lambda)^2}{2}i \int \frac{d^3k}{(2\pi)^3} \frac{1}{4 E_1 E_2} \\
&\times \Biggl( \frac{1}{i\omega - E_1 - E_2} -\frac{1}{i\omega + E_1 + E_2} \Biggr), 
\end{split}
\end{align}

is the vacuum part equal to Eq.~(\ref{fishDE}) in the deep euclidean region. 
Notice that we will use the notation $B^T(q)$ for the total fish diagram, including both vacuum and the thermal corrections, whereas $B_T(q)$ will refer only to thermal corrections to the fish
\begin{eqnarray} \label{parteT}
&&B_T(q) =  \frac{(-i\lambda)^2}{2}i \int \frac{d^3k}{(2\pi)^3} \frac{1}{4 E_1 E_2} \nonumber \\
&\times&\Biggl[ (n_1+n_2) \Biggl( \frac{1}{i\omega - E_1 - E_2} -\frac{1}{i\omega + E_1 + E_2} \Biggr) \nonumber \\  
&-&(n_1-n_2)\Biggl( \frac{1}{i\omega - E_1 + E_2} -\frac{1}{i\omega + E_1 - E_2} \Biggr)\Biggr], \end{eqnarray}
is the temperature dependent part. Since the contribution to the renormalon comes from the deep euclidean region, we calculate Eq.~(\ref{parteT}) in the limit  $-q^2\gg m^2$, obtaining
\begin{eqnarray}
B_T(q) &\approx& \frac{i \lambda^2}{4 \pi^2 q^2} \int_0^\infty{dk} \frac{k^2}{\sqrt{k^2+m^2}}  \frac{1}{e^{\sqrt{k^2+m^2}/T}-1} \nonumber \\
&\equiv& \frac{i \lambda^2}{4 \pi^2 q^2} h(\beta),
\end{eqnarray}
with
\begin{equation}
h(\beta)= \int_0^\infty{dk} \frac{k^2}{\sqrt{k^2+m^2}}  \frac{1}{e^{\sqrt{k^2+m^2}/T}-1}     
\end{equation}
Hence the contribution for one bubble is     
\begin{equation}
B^{T}(q) \approx \frac{-i\lambda^2}{32\pi ^2} \left( \log(q^2/\mu^2) -\frac{8}{q^2} h(\beta) \right).
\label{bubbleT}
\end{equation}
Now we have to insert this temperature dependent fish diagram term in the chain of bubbles that define the renormalon-type diagram. We find
\begin{eqnarray}
R_{T,k}&=&\frac{1}{(-i\lambda)^{k-2}} \int \frac{d^4q}{(2\pi)^4} i D(p+q)\left( \frac{-i \lambda^2}{32 \pi^2}\right)^{k-1} \nonumber \\
&\times& \left( \log(q^2/\mu^2) -\frac{8}{q^2} h(\beta) \right)^{k-1},
\end{eqnarray}
As we already mentioned, we proceed to expand the propagator $D(p+q)$ in powers of $1/q^2$ neglecting the first two terms that give rise to divergent integrals. We have then to calculate
\begin{eqnarray}
&&R_{T,k}=\frac{1}{(-i\lambda)^{k-2}} \int \frac{d^4q}{(2\pi)^4} \left[\frac{m^4}{q^6}-\frac{m^6}{q^8}+ \cdots \right] \nonumber \\
&\times& \left( \frac{-i \lambda^2}{32 \pi^2}\right)^{k-1} \left( \log(q^2/\mu^2) -\frac{8}{q^2} h(\beta) \right)^{k-1}.
\end{eqnarray}
Using the binomial theorem
\begin{equation}
    (A+B)^N=A^N+N \cdot A^{N-1} B + \cdots,
\end{equation}
we get
\begin{align}
\begin{split}
R_{T,k}=-i \frac{\lambda^k}{(32 \pi^2)^{k-1}} \int  \frac{d^4q}{(2\pi)^4} \left[\frac{m^4}{q^6}-\frac{m^6}{q^8}+...\right] \\
\times  \Biggl[\log(q^2/\mu^2)^{k-1} -(k-1)\frac{8}{q^2} h(\beta) \\\times\log(q^2/\mu^2)^{k-2} + ... \Biggr] \\ 
\end{split}
\end{align} 
Notice that the vacuum leading term as well as the next to leading order (NLO) vacuum term  can be extracted from the first two terms inside the first square parenthesis, together with the first term of the second square parenthesis in the previous equation. The leading term in the magnetic field strength is obtained by multiplying the first term of the first square parenthesis with the second term of the second square parenthesis in the above equation. The next terms are sub-leading.

In this way, following the same procedure of section (\ref{secRenVacio}) and performing the angular integrals we find

\begin{eqnarray}
&&R_{T,k} = -i 4m^4 \left( \frac{\lambda}{32\pi ^2}  \right) ^k \int dq \Biggl[\frac{\left(\log (q^2/\mu^2)\right)^{k-1}}{q^3}  \nonumber \\
&-&\frac{m^2}{q^5}\log (q^2/\mu^2)^{k-1} \nonumber \\
&-& \frac{(k-1)8h(\beta) (\log(q^2/\mu^2))^{k-2}}{q^5} + ... \Biggr]  \text{.} \nonumber \\
\end{eqnarray}

Using the change of variable  $q=\mu e^t$, $dq=\mu e^tdt$, 

\begin{eqnarray}
&&R_{T,k} = -i \left( \frac{\lambda}{32\pi ^2} \right)^k \frac{4m^4}{\mu^2} \int dt \Biggl[e^{-2t}(2t)^{k-1} \nonumber \\
&-& \frac{m^2}{\mu^2}e^{-4t} (2t)^{k-1} - \frac{(k-1)8h(\beta) e^{-4t}(2t)^{k-2}}{ \mu^2} + ... \Biggr]  \text{,} \nonumber \\
\end{eqnarray}
we see the appearance of the Gamma function in both terms. Using the definition of $\widetilde{m}$, we have 

\begin{eqnarray}
R_{T,k} &=& -i \widetilde{m}^2 \left( \frac{\lambda}{32\pi ^2}  \right)^k \Gamma(k) +i \left( \frac{\lambda}{64\pi ^2} \right)^k \frac{m^2}{\mu^2}\widetilde{m}^2 \Gamma (k) \nonumber \\
 &+&2i \frac{\widetilde{m}^2}{\mu^2} \left( \frac{\lambda}{64\pi ^2}  \right) ^k 8h(\beta) \Gamma(k) + ... \hspace{1mm}  \text{,}
\end{eqnarray}
where we have used also the property $\Gamma(z+1)=z\Gamma(z)$.
Now we can find the Borel transform $B[b]$ of $\Sigma R_{T,k}$,
\begin{eqnarray}
B[b]&=&\sum_k \left( \frac{R_{T,k}}{\lambda ^k}  \right) \frac{b^{k-1}}{(k-1)!} \text{,}\nonumber \\
&=& - i\widetilde{m}^2 \sum_k \left( \frac{1}{32\pi ^2}  \right) ^k b^{k-1} \nonumber \\
&+&\left(\frac{i m^2}{\mu^2}\widetilde{m}^2+\frac{i16 h(\beta) \widetilde{m}^2}{ \mu^2 }\right) \sum_k \left( \frac{1}{64\pi ^2}  \right)^k b^{k-1} + ... . \nonumber \\
\end{eqnarray}
These sums correspond to well known geometrical series, obtaining then our final result
\begin{eqnarray}
B[b] &=& \frac{-i \widetilde{m}^2}{b-32\pi ^2} \nonumber \\
&+&\left(i \frac{m^2}{\mu^2}\widetilde{m}^2+ \frac{i16 h(\beta) \widetilde{m}^2}{ \mu^2}\right) \frac{1}{b-64\pi ^2} + ...  \text{.} 
\end{eqnarray}
This result coincides with that reported in \cite{Loewe}. 
Although the location of the poles does not depend on  temperature, the residues of these poles get a temperature dependence, and this behavior depends on the function $h(\beta)$. For details, see the appendix \ref{apendice}.

%%%%%%%%%%%%%%%%%%%%%%%%%%%%%%%%%%%%%%%%%%%%%%%%%%%%%%%%%%%%%%%%%
\section{THERMOMAGNETIC RENORMALONS}\label{thermomagneticrenormalons}
In this section we are going to calculate the renormalon contribution at finite temperature taking also into account the presence of a weak external magnetic field.  The expression for the diagram to be calculated now is given by 
\begin{align}
R_{B,T,k}=\frac{1}{(-i\lambda)^{k-2}} \int \frac{d^4q}{(2\pi)^4} i D(p+q)[B^{(T,B)}(q)]^{k-1},
\end{align}
where $(T,B)$ refers to finite temperature and weak magnetic field effects. First, let us consider one bubble
\begin{eqnarray}
B^{(T,B)}(q)	&=&\frac{(-i\lambda)^2}{2} iT\sum_n\int\frac{d^3k}{(2\pi)^3}\; i\widetilde{D}^{(T,B)}(k)\; \nonumber \\
&\times& i\widetilde{D}^{(T,B)}(k-q)
\label{uno}
\end{eqnarray}
where $i\widetilde{D}^{(T,B)}(k)$ is is the finite temperature propagator up to order $(eB)^2$ in the magnetic field, defined as
\begin{align}
i\widetilde{D}^{(T,B)}(k)  &\equiv iD^{T}(k)+iD^{(T,B)}(k)     
\end{align}
with 
\begin{eqnarray}
    iD^{T}(k)&=&-\frac{i}{\omega_n^2+\vec{k}^2+m^2} \hspace{1cm}\text{and} \nonumber \\
iD^{(T,B)}(k)&=&\frac{i|eB|^2}{(\omega_n^2+\vec{k}^2+m^2)^3} \nonumber \\
&-&\frac{2i|eB|^2\;k_{\perp}^2}{(\omega_n^2+\vec{k}^2+m^2)^4}.
\end{eqnarray}
Therefore, using this notation Eq.(\ref{uno}) becomes
\begin{align}
\begin{split}
B^{(T,B)}(q)	=&\frac{(-i\lambda)^2}{2} iT\sum_n\int\frac{d^3k}{(2\pi)^3}\\ &\times(iD^{T}(k)+iD^{(T,B)}(k)) \\
&\times (iD^{T}(k-q)+iD^{(T,B)}(k-q)),  
\label{dos}
\end{split}
\end{align}
Note that the previous multiplication will produce terms of order greater than $(eB)^2$. If we restrict ourselves up to order $(eB)^2$, we obtain 
\begin{equation}
    B^{(T,B)}(q)	= D_1(k,q)+ D_2(k,q) + D_3(k,q),
\end{equation}
with 
\begin{align}
D_1(k,q)	&=\frac{(-i\lambda)^2}{2} iT \sum_n\int\frac{d^3k}{(2\pi)^3}iD^{T}(k)\;iD^{T}(k-q),\\
D_2(k,q)	&=\frac{(-i\lambda)^2}{2} iT \sum_n\int\frac{d^3k}{(2\pi)^3}iD^{T}(k)\;iD^{(T,B)}(k-q),\\
D_3(k,q)	&=\frac{(-i\lambda)^2}{2} iT \sum_n\int\frac{d^3k}{(2\pi)^3}iD^{(T,B)}(k)\;iD^{T}(k-q).
\end{align}
It is straightforward to prove that $D_2(k,q)=D_3(k,q)$.
Notice that $D_1(k,q)$ is the bubble with only temperature corrections obtained in the previous section (Eq.(\ref{fishT})), hence 
\begin{align}
D_1(k,q)	&=\frac{-i\lambda^2}{32\pi^2}\left[\log\left(\frac{q^2}{\mu^2}\right)-\frac{8}{q^2}h(\beta)\right].
\end{align}
Now we have to calculate $D_3(k,q)$. Since we know that $D_2(k,q)=D_3(k,q)$, we have 
\begin{align}
D_3(k,q)=&\frac{(-i\lambda)^2}{2} iT \sum_n\int\frac{d^3k}{(2\pi)^3}iD^{(T,B)}(k)\;iD^{T}(k-q) \nonumber \\
=& \frac{(-i\lambda)^2}{2} iT \sum_n\int\frac{d^3k}{(2\pi)^3} \nonumber\\
&\times \Biggl[ \frac{i|eB|^2}{(\omega_n^2+\vec{k}^2+m^2)^3} -\frac{2i|eB|^2\;k_{\perp}^2}{(\omega_n^2+\vec{k}^2+m^2)^4}  \Biggr] \nonumber\\
&\times \Biggl[ -\frac{i}{(\omega_n-\omega)^2+(\vec k-\vec q)^2+m^2} \Biggr], \nonumber \\
\equiv &D_{3.1}(k,q) + D_{3.2}(k,q)
\end{align}
with
\begin{align}
\begin{split}
D_{3.1}(k,q)=&  \frac{(-i\lambda)^2 i T}{2} \sum_n\int\frac{d^3k}{(2\pi)^3}  \frac{|eB|^2}{(\omega_n^2+\vec{k}^2+m^2)^3} \\&\times \Biggl[ \frac{1}{(\omega_n-\omega)^2+(\vec k-\vec q)^2+m^2} \Biggr], 
\end{split}
\end{align}
and
\begin{align}
\begin{split}
D_{3.2}(k,q)=&\frac{(-i\lambda)^2 i T}{2}\sum_n\int\frac{d^3k}{(2\pi)^3} \frac{2|eB|^2\;k_{\perp}^2}{(\omega_n^2+\vec{k}^2+m^2)^4} \\
&\times \Biggl[ \frac{-1}{(\omega_n-\omega)^2+(\vec k-\vec q)^2+m^2} \Biggr]. 
\end{split}
\end{align}
Let's first calculate $D_{3.1}(k,q)$, 
\begin{eqnarray}
 &&D_{3.1}(k,q)=  \frac{(-i\lambda)^2}{2\cdot 2!}\left(\frac{\partial}{\partial \tilde{m}^2}\right)^2 |eB|^2\;i T\sum_{n}\int\frac{d^3k}{(2\pi)^3} \nonumber \\
 &\times& \Biggl[ \frac{1}{\omega_n^2+\vec{k}^2+\tilde{m}^2} \Biggr] 
 \Biggl[ \frac{1}{(\omega_n-\omega)^2+(\vec k-\vec q)^2+m^2} \Biggr], 
\end{eqnarray}
where we have used
\begin{equation}
    \frac{1}{2!}\left(\frac{\partial}{\partial \tilde m^2}\right)^2\frac1{k^2-\tilde m^2}	=\frac1{(k^2-\tilde m^2)^3}
\end{equation}
let us note that we have again the expression from the previous section (Eq. (\ref{26})), with the difference that we have to derive with respect to $\tilde{m}^2$ twice. In this way, we obtain
\begin{eqnarray}
&&D_{3.1}(k,q)	=\frac{|eB|^2(-i\lambda)^2 \;i}{2\cdot 2!}\nonumber \\
&&\times\left[\frac{2!}{32\pi^2 m^2 q^2}-\frac1{2\pi^2 q^2}\left(\frac{\partial}{\partial \tilde m}\right)^2 h(\beta) \right].
\end{eqnarray}
Note that the derivatives in $\tilde{m}$ have not been calculated because, as in the previous section, the function  $h(\beta)$ needs to be analyzed separately for the low and high temperature cases. In a similar fashion we calculate $D_{3.2}(k,q)$,
\begin{eqnarray}
 &&D_{3.2}(k,q)=  \frac{(-i\lambda)^2}{3!}\left(\frac{\partial}{\partial \tilde{m}^2}\right)^3 |eB|^2\;i T\sum_{n}\int\frac{d^3k}{(2\pi)^3} \nonumber \\
 &\times& \Biggl[ \frac{k_{\perp}^2}{\omega_n^2+\vec{k}^2+\tilde{m}^2} \Biggr] 
 \Biggl[ \frac{-1}{(\omega_n-\omega)^2+(\vec k-\vec q)^2+m^2} \Biggr], 
\end{eqnarray}
where we have  used
\begin{equation}
    \frac{1}{3!}\left(\frac{\partial}{\partial \tilde m^2}\right)^3\frac1{k^2-\tilde m^2}=\frac1{(k^2-\tilde m^2)^4},
\end{equation}
let us note that again we have the expression from the previous section (Eq. (\ref{26})), with the difference that we have to derive three times with respect to $\tilde{m}^2$. Hence, we obtain
\begin{eqnarray}
&&D_{3.2}(k,q)	=\frac{|eB|^2(-i\lambda)^2\; i}{2\cdot 3!}\nonumber
\\&&\times\left[\frac{-3!}{96\pi^2 m^2 q^2}-\frac{2}{6\pi^2 q^2} \left(\frac{\partial}{\partial \tilde m^2}\right)^3 k^2h(\beta) \right].
\end{eqnarray}Taking into account the results $D_1(k,q)$, $D_2(k,q)$ and $D_3(k,q)$, we obtain for $B^{(T,B)}(q)$ 
\begin{multline}
B^{(T,B)}(q)	=\frac{(i\lambda)^2}{2} \left[\frac{i}{16\pi^2} \log \left(\frac{-q^2}{\mu^2}\right)-\frac{8i}{16\pi^2q^2}h(\beta) \right.\\
		+2|eB|^2 i \left\{\frac{1}{48\pi^2 m^2 q^2} - \frac{1}{2\pi^2 q^2}\frac{1}{2!}\left(\frac{\partial}{\partial \tilde m^2}\right)^2 h(\beta) \right.\\
		\left.\left.-\frac{1}{2\pi^2 q^2}\frac{1}{3!} \left(\frac{\partial}{\partial \tilde m^2}\right)^3 h(\beta) \cdot k^2\right\}\right].
\end{multline}
As mentioned in the previous section, the function $h(\beta)$ cannot be calculated in a closed analytic form, but it can be expressed analytically for  the cases of high and low temperature. Therefore, before calculating the renormalon diagram, we will calculate $B^{(T,B)}(q)$ for the case of high and low temperature. For the high temperature case, we calculate the corresponding derivatives using the appendix \ref{apendice} obtaining
\begin{eqnarray}
B_{\text{HT}}(q)	&=&-\frac{i\lambda^2}{32\pi^2}\Biggl[ \log\left( \frac{q^2}{\mu^2}\right)+\frac{1}{q^2}\Biggl(-\frac{4\pi^2 T^2}{3} + 2 \pi m T \nonumber \\
&& -2 m^2 \Biggl( \frac{1}{2} + \log\left(\frac{4 \pi T}{m}\right) -\gamma  \Biggr)  - \frac{m^4}{T^2}\zeta(3)\nonumber \\
&+&\frac{m^6}{8 \pi^4 T^4} \zeta(5)    +|eB|^2 \Biggl[\frac{7}{32 m^2}-\frac{\pi T}{m^3} \nonumber \\
&+& \frac{1}{2 \pi^2 T^2}\zeta(3) - \frac{3m^2}{16\pi^4 T^4} \zeta(5) \Biggr] \Biggr)  \Biggr].
\end{eqnarray}
For the low temperature case, we have
\begin{eqnarray}
&&B_{\text{LT}}(q)	=-\frac{i\lambda^2}{32\pi^2}\Biggl[ \log\left( \frac{q^2}{\mu^2}\right)+\frac{1}{q^2}\Biggl(-8\left(\frac{T^3m\pi}{2}\right)^{1/2} \nonumber \\
&\times &\plog{3/2}\left(e^{-m/T} \right)+\frac{2|eB|^2}{3m^2} 
		-32 |eB|^2\Biggl[ \frac{1}{32m^3}\left(\frac{\pi}{2m T}\right)^{1/2}\nonumber \\
		&\times&\left[4m^2 \plog{-1/2}\left(e^{-m/T} \right)-3T^2 \plog{3/2}\left(e^{-m/T} \right) \right] 
		\nonumber \\
		&+&\frac{1}{384}\left(\frac{\pi}{2m^9 T}\right)^{1/2} \Biggl[ 3T\Biggl( 4m^2 \plog{1/2}\left(e^{-m/T}\right)   
		\nonumber \\
		&+&5T \left\{2m\; \plog{3/2}\left(e^{-m/T} \right)+T\;\plog{5/2}\left(e^{-m/T} \right) \right\} \Biggr)  \nonumber \\
		&-&8m^3 \plog{-1/2}\left(e^{-m/T} \right) \Biggr] \Biggr] \Biggr)  \Biggr].
\end{eqnarray}
At this point, we have to calculate the renormalon-type diagram, both for  high and low temperature regions, in the presence of a weak magnetic field.

\subsection{Magnetic Renormalons in the High Temperature region}
Following the same procedure of the previous section, we have to calculate the renormalon-type diagram
\begin{align}
R^\text{HT}_{B,T,k}=\frac{1}{(-i\lambda)^{k-2}} \int \frac{d^4q}{(2\pi)^4} i D(p+q)[B_{{\text{HT}}}^{(T,B)}(q)]^{k-1},
\end{align}
obtaining
\begin{eqnarray} 
R^\text{HT}_{B,T,k} &=& -i \widetilde{m}^2 \left( \frac{\lambda}{32\pi ^2}  \right)^k \Gamma(k) +i \left( \frac{\lambda}{64\pi ^2} \right)^k \frac{m^2}{\mu^2}\widetilde{m}^2 \Gamma (k) \nonumber \\
 &-&2i \frac{\widetilde{m}^2}{\mu^2} \left( \frac{\lambda}{64\pi ^2}  \right)^k \Gamma(k) \Biggl(-\frac{4\pi^2 T^2}{3} + 2 \pi m T \nonumber \\
&& -2 m^2 \Biggl( \frac{1}{2} + \log\left(\frac{4 \pi T}{m}\right) -\gamma  \Biggr)  - \frac{m^4}{T^2}\zeta(3)\nonumber \\
&+&\frac{m^6}{8 \pi^4 T^4} \zeta(5)    +|eB|^2 \Biggl[\frac{7}{32 m^2}-\frac{\pi T}{m^3} \nonumber \\
&+& \frac{1}{2 \pi^2 T^2}\zeta(3) - \frac{3m^2}{16\pi^4 T^4} \zeta(5) \Biggr] \Biggr)  + ...   \nonumber \\
\end{eqnarray}
This can be written as
\begin{eqnarray}
R^\text{HT}_{B,T,k}&\equiv& -i \widetilde{m}^2 \left( \frac{\lambda}{32\pi ^2}  \right)^k \Gamma(k) +i \left( \frac{\lambda}{64\pi ^2} \right)^k \frac{m^2}{\mu^2}\widetilde{m}^2 \Gamma (k) \nonumber \\
 &-&2i \frac{\widetilde{m}^2}{\mu^2} \left( \frac{\lambda}{64\pi ^2}  \right)^k \Gamma(k) F_{\text{HT}} + ... \hspace{1mm}  \text{,}
\end{eqnarray}
where
\begin{eqnarray}
F_{\text{HT}}&=& -\frac{4\pi^2 T^2}{3} + 2 \pi m T \nonumber \\
&& -2 m^2 \Biggl( \frac{1}{2} + \log\left(\frac{4 \pi T}{m}\right) -\gamma  \Biggr)  - \frac{m^4}{T^2}\zeta(3)\nonumber \\
&+&\frac{m^6}{8 \pi^4 T^4} \zeta(5)    +|eB|^2 \Biggl[\frac{7}{32 m^2}-\frac{\pi T}{m^3} \nonumber \\
&+& \frac{1}{2 \pi^2 T^2}\zeta(3) - \frac{3m^2}{16\pi^4 T^4} \zeta(5) \Biggr]    \end{eqnarray}
\begin{figure}
\begin{center}
\includegraphics[scale=0.38]{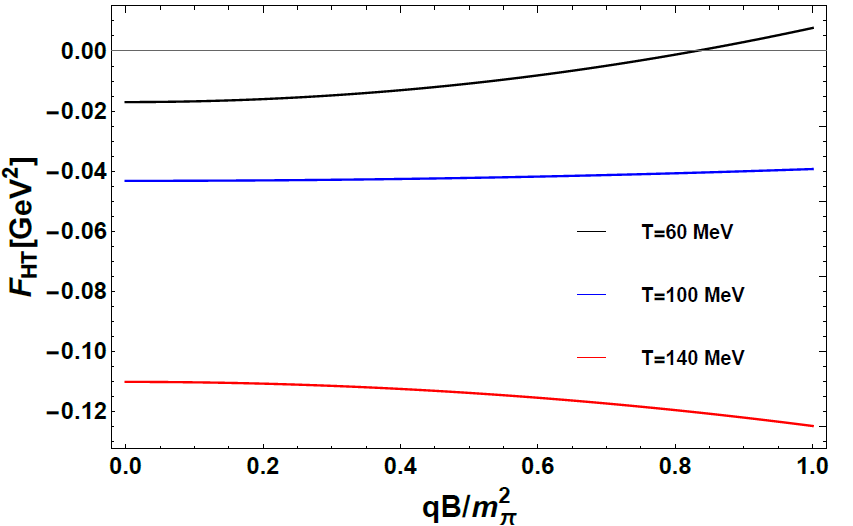}
\caption{$F_{\text{HT}}$ vs. $T$.}
\label{residuoHT}
\end{center}
\end{figure}
Now we can find the Borel transform $B[b]$
\begin{equation}
B[b] = \frac{-i \widetilde{m}^2}{b-32\pi ^2} +\Biggl[i \frac{m^2}{\mu^2}\widetilde{m}^2- \frac{2i \widetilde{m}^2}{ \mu^2}F_{\text{HT}} \Biggr] \frac{1}{b-64\pi ^2} + ...  \text{.} 
\end{equation}

As we mentioned, the location of the poles does not depend on temperature nor on the magnetic field. However, from Fig. \ref{residuoHT} we notice a competition between temperature and the strength of the magnetic field for the evolution of the residue.  For $T=60$ MeV the curve grows with the magnetic field but its concavity is low. When we start increasing temperature, this concavity becomes smaller until a value of temperature where the sign of the concavity changes, having then a decreasing behavior. This indicates that temperature becomes dominant with respect to the magnetic field.  

\subsection{Magnetic Renormalons in the Low Temperature region}

\begin{align}
R^\text{LT}_{B,T,k}=\frac{1}{(-i\lambda)^{k-2}} \int \frac{d^4q}{(2\pi)^4} i D(p+q)[B_{{\text{LT}}}^{(T,B)}(q)]^{k-1},
\end{align}
Again, we calculate in the same fashion as previous section. We obtain
\begin{align} 
R^\text{LT}_{B,T,k} =& -i \widetilde{m}^2 \left( \frac{\lambda}{32\pi ^2}  \right)^k \Gamma(k) +i \left( \frac{\lambda}{64\pi ^2} \right)^k \frac{m^2}{\mu^2}\widetilde{m}^2 \Gamma (k)
\nonumber \\
&-2i \frac{\widetilde{m}^2}{\mu^2} \left( \frac{\lambda}{64\pi ^2}  \right)^k \Gamma(k) \nonumber\\
&\times \Biggl(-8\left(\frac{T^3m\pi}{2}\right)^{1/2} \plog{3/2}\left(e^{-m/T} \right) \nonumber \\
&+\frac{2|eB|^2}{3m^2} 
		-32 |eB|^2\Biggl[ \frac{1}{32m^3}\left(\frac{\pi}{2m T}\right)^{1/2} \nonumber \\
&\times\bigl[4m^2 \plog{-1/2}\left(e^{-m/T} \right)	-3T^2 \plog{3/2}\left(e^{-m/T} \right) \bigr]
		\nonumber \\
&+\frac{1}{384}\left(\frac{\pi}{2m^9 T}\right)^{1/2} \biggl[ 3T\Biggl( 4m^2 \plog{1/2}\left(e^{-m/T}\right) 
	 \nonumber \\
&+5T \biggl\{2m\; \plog{3/2}\left(e^{-m/T} \right) +T\;\plog{5/2}\left(e^{-m/T} \right) \biggr\} \Biggr) 
	 \nonumber \\
&-8m^3 \plog{-1/2}\left(e^{-m/T} \right) \biggr] \Biggr] \Biggr) + ... 
\end{align}
\noindent It is convenient to write this, as we did in the high temperature case, as
\begin{align}
\begin{split}
R^\text{LT}_{B,T,k}	\equiv&	-i \widetilde{m}^2 \left( \frac{\lambda}{32\pi ^2}  \right)^k \Gamma(k) +i \left( \frac{\lambda}{64\pi ^2} \right)^k \frac{m^2}{\mu^2}\widetilde{m}^2 \Gamma (k) \\
&-2i \frac{\widetilde{m}^2}{\mu^2} \left( \frac{\lambda}{64\pi ^2}  \right)^k \Gamma(k) F_{\text{LT}} + ... \hspace{1mm}  \text{,}
\end{split}
\end{align}
where
\begin{eqnarray}
&&F_{\text{LT}}=-8\left(\frac{T^3m\pi}{2}\right)^{1/2} \plog{3/2}\left(e^{-m/T} \right) \nonumber \\
&&+\frac{2|eB|^2}{3m^2} 
		-32 |eB|^2\Biggl[ \frac{1}{32m^3}\left(\frac{\pi}{2m T}\right)^{1/2} \nonumber \\
		&\times&\bigl[4m^2 \plog{-1/2}\left(e^{-m/T} \right)	-3T^2 \plog{3/2}\left(e^{-m/T} \right) \bigr]
		\nonumber \\
		&&+\frac{1}{384}\left(\frac{\pi}{2m^9 T}\right)^{1/2} \biggl[ 3T\Biggl( 4m^2 \plog{1/2}\left(e^{-m/T}\right) 
	 \nonumber \\
	 &&+5T \biggl\{2m\; \plog{3/2}\left(e^{-m/T} \right) +T\;\plog{5/2}\left(e^{-m/T} \right) \biggr\} \Biggr) 
	 \nonumber \\
	 &&-8m^3 \plog{-1/2}\left(e^{-m/T} \right) \biggr] \Biggr].
\end{eqnarray}

Now we can find the Borel transform $B[b]$
\begin{equation}
B[b] = \frac{-i \widetilde{m}^2}{b-32\pi ^2} +\Biggl[i \frac{m^2}{\mu^2}\widetilde{m}^2- \frac{2i \widetilde{m}^2}{ \mu^2} F_{\text{LT}}\Biggr] \frac{1}{b-64\pi ^2} + ...  \text{.} 
\end{equation}

\begin{figure}[h]
\includegraphics[width=85mm]{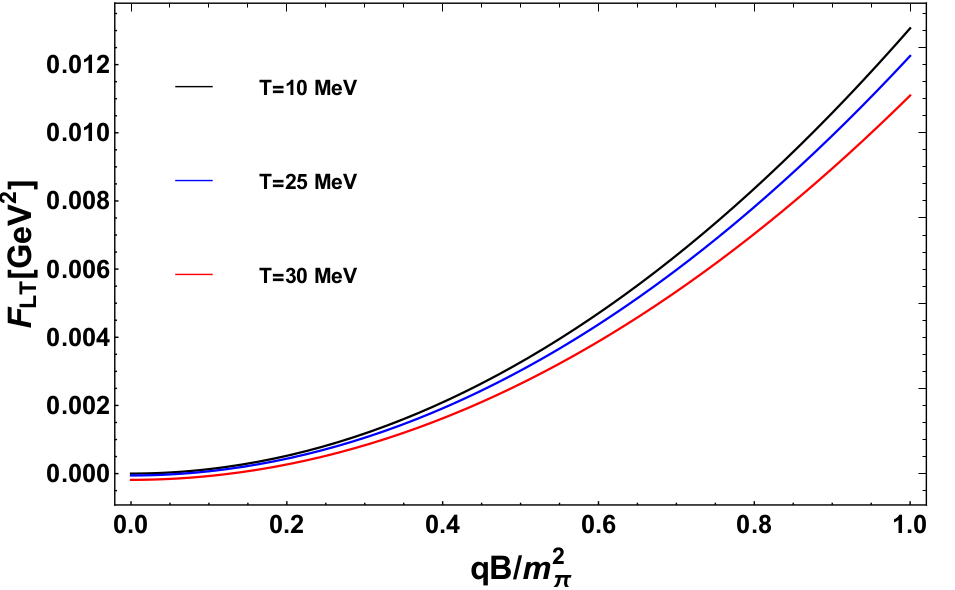} 
\caption{$F_{\text{LT}}$ vs. $T$.}
\label{residuoLT}
\end{figure}

Again, we see that  for low temperature and weak magnetic field, the location of the renormalons does not depend on temperature or on the magnetic field. As we can see in  Fig. \ref{residuoLT}, the low temperature region,  we have a growing evolution of the residue with the magnetic field. The curvature is bigger than in the high temperature region. This is reasonable,since we are in the low temperature region where the magnetic field dominates although we already notice that for higher values of temperature the tendency of the residue is to diminish. 

\section{CONCLUSIONS}\label{conclusions}

Using the imaginary time formalism for dealing with temperature effects and the Schwinger propagator for handling the external magnetic field, we have analyzed the influence of temperature and of a weak external magnetic field on the renormalon diagram in the theory $\lambda \phi^4$.

Initially we considered only temperature effects, finding the same results obtained in the frame o thermo-field dynamics \cite{Loewe}, being this a formalism in real time. This confirms again the validity of both formalisms.

In the case of the thermo-magnetic corrections to our renormalon type of diagram,  we get a non-analytic closed expression. However, for the low and high temperature regions it is possible to have analytic expressions. The agreement between our analytic expression in the high temperature regions and the numerical results is amazing, being valid for  a wide temperature region. Something interesting is the competition we found, in the calculation of the residue, between the magnetic field and temperature. As we said, in the low temperature region, and low magnetic field strength, the residue grows. As soon as temperature starts to increase, this growing behavior becomes less pronounced. We also found, now in the high temperature region, that, for example, for $T = 60$ MeV, we still have a growing behavior but with much less concavity. In the case of really high temperature, the residue diminishes with the magnetic field. It is interesting to point out that this scenario was also found in the behavior of the $\pi-\pi$ scattering lengths \cite{pi1,pi2} where we have exactly the same competition between temperature and magnetic field strength.

%%%%%%%%%%%%%%%%%%%%%%%%%%%%%%%%%%%%%%%%%%%%%%%%%%%%%%%%%%%%%%%%%
\section*{Acknowledgements}
M. Loewe and R. Zamora acknowledge support from   ANID/CONICYT FONDECYT Regular (Chile) under Grant No. 1200483. M.L.  acknowledges support from Fondecyt under grants No. 1190192 and No. 1170107. ML acknowledges also support from Anid/Pia/Basal (Chile) under grant No. FB082

%%%%%%%%%%%%%%%%%%%%%%%%%%%%%%%%%%%%%%%%%%%%%%%%%%%%%%%%%%%%%%%%%

\begin{appendix}
\section{}\label{apendice}
In this appendix we are going to analyze the function $h(\beta)$. Being this function non analytic, we are going to study separately its behavior for the  high and low temperature cases.

We will start with the high temperature case $h_{HT}(\beta)$. We have
\begin{align}
h(\beta)	&=\int_0^\infty \frac{k^2\;dk}{\sqrt{k^2+m^2}}\frac{1}{e^{\sqrt{k^2+m^2}/T}-1},
\end{align}
Defining $m/T=y$ and $k/T=x$, we get
\begin{equation}
h(\beta)= 
T^2 \int_0^\infty \frac{dx\; x^2}{\sqrt{x^2+y^2}}\frac{1}{e^{\sqrt{x^2+y^2}}-1},
\end{equation}
The last expression can be related to the functions presented in the appendix of \cite{Kapusta}. We have $h(\beta)=T^2\Gamma(3)h_3(y)=2T^2h_3(y)$. To find $h_3(y)$, we use the relation
\begin{equation}
\frac{dh_3(y)}{dy}=\frac{-yh_1(y)}{2},
\end{equation}
where
\begin{eqnarray}
h^{HT}_1(y)&=&\frac{\pi}{2y}+\frac12 \log\left(\frac{y}{4\pi}\right)+\frac12 \gamma_E  -\frac{1}{4}\zeta(3) \left(\frac{y}{2 \pi}\right)^2\nonumber \\
&+& \frac{3}{16}\zeta(5) \left(\frac{y}{2 \pi}\right)^4 +\cdots ,    
\end{eqnarray}
Therefore, we obtain
\begin{eqnarray}
\frac{dh^{HT}_3(y)}{dy}	&=&-\frac\pi4 -\frac{y}{4} \log\left(\frac{y}{4\pi}\right)-\frac{y}{4} \gamma_E +\frac{y}{8}\zeta(3) \left(\frac{y}{2 \pi}\right)^2 \nonumber \\
&-& \frac{3y}{32}\zeta(5) \left(\frac{y}{2 \pi}\right)^4 +\cdots,
\end{eqnarray}
Solving the previous equation
\begin{eqnarray}
h^{HT}_3(y)	&=&-\frac{\pi y}{4}+\frac{y^2}{8}\left[\frac12 +\log\left(\frac{4\pi}{y}\right)-\gamma\right]\nonumber \\
&+&\frac{y^4}{128 \pi^2}\zeta(3)-\frac{y^6}{1024 \pi^2}\zeta(5) +C_1 +\cdots.
\end{eqnarray}
In order to find the constant of integration ($C_1$), we calculate
\begin{align}
h_3(0)	&=\frac{1}{\Gamma(3)}\int_0^\infty dx\frac{x^2}{x}\frac{1}{e^x-1}=\frac{1}{\Gamma(3)}\frac{\pi^2}{6}=\frac{\pi^2}{12},\\
&\Rightarrow C_1=\frac{\pi^2}{12},
\end{align}
therefore,
\begin{eqnarray}
   h^{HT}_3(y)&=&\frac{\pi^2}{12}-\frac{\pi y}{4}+\frac{y^2}{8}\left[ \frac12+\log\left(\frac{4\pi}{y}\right)-\gamma \right] \nonumber \\
   &+&\frac{y^4}{128 \pi^2}\zeta(3)-\frac{y^6}{1024 \pi^2}\zeta(5)+\cdots.
\end{eqnarray}
Finally,  for the high temperature case we obtain 
\begin{eqnarray}
h_{HT}(\beta)&=&2T^2\;h^{HT}_3(y)=2T^2\;h^{HT}_3\left(\frac mT\right) \nonumber\\
		&=&2T^2\Biggl(\frac{\pi^2}{12}-\frac{\pi m}{4T} +\frac{m^2}{8T^2}\left[ \frac12+\log\left(\frac{4\pi T}{m}\right)-\gamma \right] \nonumber \\ &+&\frac{m^4}{128 \pi^2 T^4}\zeta(3) 
		-\frac{m^6}{1024 \pi^4 T^6}\zeta(5)+\cdots \Biggr).
\end{eqnarray}

Now we will analyze the behavior of $h(\beta)$ in the low temperature regime. We have
\begin{align}
h(\beta)	&=\int_0^\infty \frac{k^2\;dk}{\sqrt{k^2+m^2}}\frac{1}{e^{\sqrt{k^2+m^2}/T}-1},
\end{align}
using $\omega^2=k^2+m^2$, we get,
\begin{align}
h(\beta)=\int_0^\infty \frac{(\omega^2-m^2)\omega\;d\omega}{\omega\sqrt{\omega^2-m^2}}\frac{1}{e^{\omega/T}-1},
\end{align}
which can be written as
\begin{equation}
h(\beta)=\int_m^\infty d\omega\;\sqrt{\omega^2-m^2} \frac{e^{-\omega/T}}{1-e^{-\omega/T}}.
\end{equation}
In the case of low temperature, the last part of the previous equation corresponds to a geometric series. Therefore,
\begin{equation}
  h_{LT}(\beta)  =\sum_{n=1}^\infty \int_m^\infty d\omega\sqrt{\omega^2-m^2} e^{-n\omega/T}.
\end{equation}
Performing the integral, we get
\begin{equation}
  h_{LT}(\beta)  =\sum_{n=1}^\infty \frac{mT}{n}K_1\left(\frac{mn}{T}\right),
\end{equation}
 Note that for low temperature
\begin{equation}
 \frac{mT}{n}K_1\left(\frac{mn}{T}\right) \\\xrightarrow[T\to 0]{}\left(\frac{T^3 m \pi}{2n^3}\right)^{1/2}e^{-mn/T},   
\end{equation}

Using the above result, we have
\begin{equation}
h_{LT}(\beta)=\sum_{n=1}^\infty \left(\frac{T^3 m \pi}{2n^3}\right)^{1/2}e^{-mn/T},	
\end{equation}
The above sum is analytical, and corresponds the Polylogarithm functions ($\text{Li}$). Finally, we obtain 
\begin{equation}
   h_{LT}(\beta) = \left(\frac{T^3 m \pi}{2}\right)^{1/2}\text{Li}_{3/2}\left(e^{-m/T}\right).
\end{equation}
Now we are going to plot the function $h(\beta)$, comparing its numerical value  with the  high temperature and low  approximations.
\begin{figure*}
\begin{center}
\includegraphics[scale=0.4]{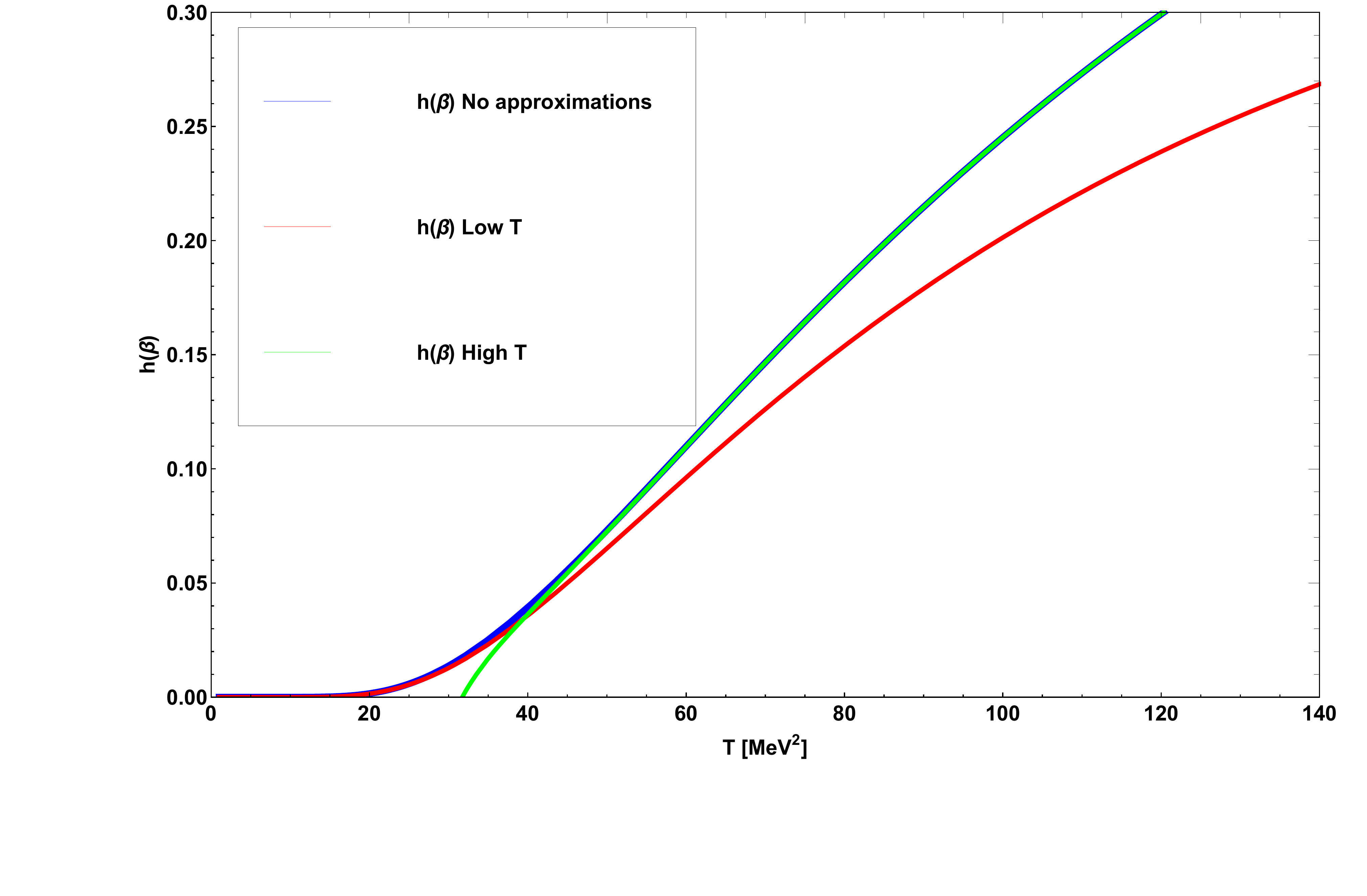} 
\caption{Comparison of the function $h(\beta)$ with the high and low temperature approximations. The blue line corresponds to the numerical evaluation of $h(\beta)$ , the red line for the case of $h(\beta)$ at low temperature and the green line for the case of $h(\beta)$ at high temperature.}
\label{comparaciondeaproximaciones}
\end{center}
\end{figure*}
We note that our approximation for the low temperature case,  up to a value of around $T\sim40$ MeV, is quite good whereas for a temperature  bigger than $T\sim40$ MeV the high temperature approximation is also excellent. Therefore, with our approximations,  we are able to cover in an analytic way the whole range of temperatures. It is important, however, to stress how sensible our analytic expressions are. For example, if we do not include the terms that involve the Riemann zeta factors in the high temperature region, the approximation turns out to be valid only for atemperature bigger than  $T\sim 140$ MeV.
\end{appendix}

%%%%%%%%%%%%%%%%%%%%%%%%%%%%%%%%%%%%%%%%%%%%%%%%%%%%%%%%%%%%%%%%%

%%%%%%%%%%%%%%%%%%%%%%%%%%%%%%%%%%%%%

\end{document}